# *In Situ* RESISTANCE MEASUREMENTS OF STRAINED CARBON NANOTUBES

S. Paulson[1], M.R. Falvo[1], N. Snider[1], A. Helser[2], T. Hudson[2], A. Seeger[2], R.M. Taylor II[2], R. Superfine[1] and S. Washburn[1,a)]

1 *Department of Physics and Astronomy, University of North Carolina-Chapel Hill*
2 *Department of Computer Science, University of North Carolina-Chapel Hill*
a) e-mail: sean@physics.unc.edu

We investigate the response of multi-walled carbon nanotubes to mechanical strain applied with an Atomic Force Microscope (AFM) probe. We find that in some samples, changes in the contact resistance dominate the measured resistance change. In others, strain large enough to fracture the tube can be applied without a significant change in the contact resistance. In this case we observe that enough force is applied to break the tube without any change in resistance until the tube fails. We have also manipulated the ends of the broken tube back in contact with each other, re-establishing a finite resistance. We observe that in this broken configuration the resistance of the sample is tunable to values 15-350 kΩ greater than prior to breaking.

Soon after their discovery by Iijima, carbon nanotubes [1](CNTs) were predicted to have interesting electrical properties[2], possibly making them suitable components for miniaturization of electronic devices [3,4] and nano-electromechanical systems (NEMS). To serve a role in NEMS, both the electrical and mechanical properties of CNTs, as well as their response to electrical and mechanical interactions must be understood. To date, several experiments have probed the electrical [5-9] and mechanical [10-13] properties. However, there have been fewer (and less controlled) efforts examining the effect of mechanical strain on the electrical properties[8], despite considerable theoretical effort to predict the effect of strain and various defects [14-18]. We present the results of our experiments using the probe of an Atomic Force Microscope (AFM) to apply a mechanical stress to multi-walled carbon nanotubes (MWNT), while monitoring the resistance *in situ*.

Two different techniques were used to produce samples, electron beam lithography, and an AFM based lithography developed here. In both cases, a solution of MWNTs in ethanol was dispensed onto a thermally oxidized silicon wafer spinning at 4000 RPM. Next, two metal leads were placed over the nanotubes. Because the samples in these experiments are subject to strain along the length of the tube, placing the leads on top of the nanotube is imperative to hold the nanotube in place. The metal clamps the MWNT down, preventing unwanted relative motion between the contact and the MWNT. We show that if relative motion does occur (if the film does not pin the MWNT), large changes in the contact resistance result, making it impossible to interpret how much the resistance of the tube is changing, if at all.

Our experiments are performed by placing the tip of an AFM on the substrate near the MWNT. The tip is attached to a silicon cantilever which, in turn, is connected to a larger silicon chip. Using the nanoManipulator software, which has been described in detail elsewhere [19], the chip is moved, dragging the tip laterally through the MWNT. The deflection of the cantilever, which acts as a spring, is monitored and is proportional to the force applied to the MWNT. A schematic is shown in Fig 1(a). Throughout the experiment, the low-bias (1mV) resistance and the (lateral and normal) force the tip applies to the MWNT are recorded. While applying stress to the samples, we have observed two types of behavior.

First, we look at a sample where the tube is of comparable thickness to the leads. Figs. 1(b-f) show the results of a series of manipulations on a 19 nm diameter MWNT under 15 nm platinum leads. A series of two manipulations was performed on this tube, with images taken after each event. Fig. 1(b) shows the nanotube before modification, the resistance is 85 kΩ. The arrow indicates the tip trajectory of the first modification, resulting in Fig. 1(c). While being pushed, the resistance increased to 220 kΩ, and remained at that value after the tip was removed. The tube underwent further manipulation, indicated by the arrow in Fig 1(c), resulting in the image in figure 1(f). The resistance during this manipulation decreased to 120 kΩ, and again remained constant even after the tip was removed.

Figures 1(d) and 1(e) show images of the end of the tube before and after manipulation. In Fig. 1(e), the trench in the metal where the tube had been is a clear indication that the tube has moved with respect to the lead, indicating that the contact resistance could be changing. Additionally, we can estimate the local strain in the MWNT as $R_c/R_t$, where $R_c$ is the radius of curvature of the tube and $R_t$ is the outer radius of the MWNT. We find the first modification leads to a maximum strain of 7%, and the second manipulation induces a maximum 11% strain. That the less strained configuration yields a higher resistance suggests that the contact resistance varies as the tube moves with respect to the metal, and this obscures any effect intrinsic to the nanotube.

To try and prevent the MWNT from moving under the contacts, we attempted pushing on the top of the tube, normal to the substrate. A force of 800 nN was applied to a 20nm diameter tube, resulting in a 5nm deformation of the tube [20]. The resistance change, if any, in this 80 kΩ sample is less than 500 Ω, the noise in the measurement.

In a second sample, with thicker metal, we observe strikingly different behavior. Figure 2 shows AFM images of a sample produced with AFM lithography. The tube is 24nm in diameter, and the Au/Pd contacts are 50 nm thick. The initial resistance of the sample is ~135 kΩ. During the modification, the resistance changes abruptly to 143 kΩ, then increases gradually to 148 kΩ as the support moves an additional 25 nm. Further pushing causes a rapid increase to 400 kΩ, and beyond this, continued motion of the tip causes the resistance to become immeasurably large ($> 10^7$ Ω). Fig. 2(d) is an AFM image that shows the tube had been severed.

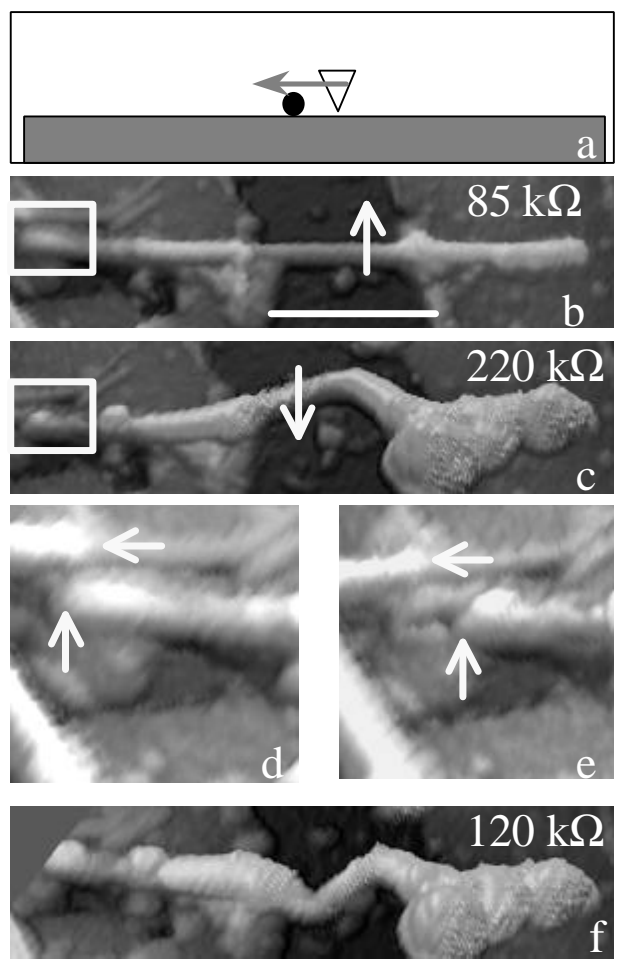

Fig. 1. (a) Schematic of experiment. The AFM tip is placed on the substrate next to the MWNT. The tip applies a constant normal force to the substrate and is pushed laterally through the tube. The lateral force and resistance are recorded while the tube is strained. (b) MWNT between electron beam defined leads. The arrow indicates the tip trajectory between (b) and (c). The scale bar is 300 nm. (c) Sample after modification depicted in (b). The arrow indicates the tip trajectory between c) and f). (d & e) Close ups of boxed areas in (b & c) respectively. The arrows show the relative motion between the tube and a stationary object, as well as the trench left by the tube. (f) Image of sample after second modification.

Looking at the force between the tip and the nanotube during the manipulation gives some insight into this behavior. Fig. 2(b) shows the lateral force applied to the tube and the resistance of the sample versus the motion of the cantilever. As the tip first contacts the nanotube, there is a linear increase in the lateral force until it reaches a maximum value, then it drops suddenly. Observing the resistance, we see it remains constant (within the noise of the measurement) as the tip contacts the tube and as the lateral force increases. As the lateral force suddenly drops, the resistance simultaneously increases to 143 kΩ. This behavior can be understood if the lateral force applied to the tube does not induce enough strain to change the resistance until the tube is stretched beyond the elastic limit, at which point it either fractures, or deforms plastically[21]. At this point the deformed tube allows the stress in the nanotube to diminish, accompanied by an immediate increase in resistance. This explanation adequately describes the result of this experiment, and it agrees with recent theoretical work, which calculates the transmission of electrons through nanotubes treated as ballistic conductors [16,22]. These calculations show that in tubes that are bent up to 90 degrees, well beyond the point at which they buckle, the low bias conduction is unchanged. We are currently pursuing current versus voltage spectroscopy experiments on single-walled nanotube samples with four probes rather than two. This will allow us to avoid the effects of contact resistance, as well as the complication of multiple conducting shells, each experiencing different strain, and breaking at different times.

Further experiments done on the same sample give us additional evidence that the initial change (to 143-148 kΩ) corresponds to the breaking MWNT. Using the AFM tip to push the two pieces back together as shown in Fig. 2(e), we find the pieces become electrically connected again. By "poking" the newly formed junction with the AFM tip, the resistance can be tuned. The values of resistance measured after the tube was broken and pushed together varied from 148 to 500 kΩ, though the resistance at a given value had as little noise as the original measurement (~2%). We speculate that this variation comes from microscopic details of the contact between the two halves, which are not yet fully understood. This suggests that nanotubes can be used as leads to study properties of nanoscopic particles, by using the two ends of the broken nanotube, or two different nanotubes near each other as a break junction with tunable contacts.

It is of interest that enough strain was applied to fracture the tube, without causing the nanotube/metal junction to fail. This indicates the strength of the interface is quite large, which contrasts tensile strength experiments with polymer nanotube composites. In these experiments the failure mechanism was "pull out" of the nanotube from the polymer matrix, rather than fracture of the MWNTs[23]. To compare the nature of these interfaces, more careful experiments must be done, as our strain was not completely uniaxial, and the substrate could have affected the results.

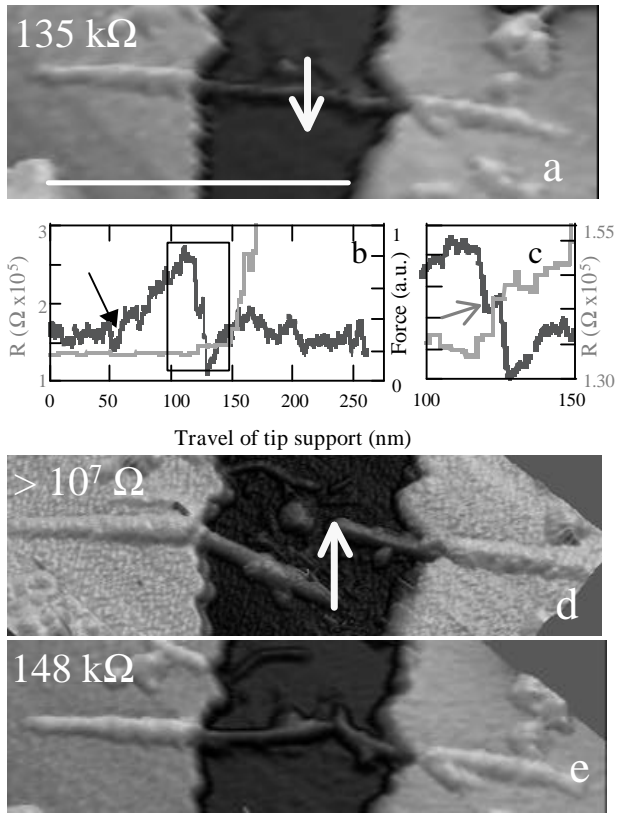

Fig. 2. (a). AFM Image of Sample produced by AFM lithography. Scale bar is 1μ. The arrow indicates the trajectory of the tip between (a) and (c). (b). Resistance (gray) and Lateral Force (black) during modification shown in a. X-axis is the distance the cantilever support moves. The arrow indicates when the tip hits the sample. (c) Zoom in of box in (b). Notice the simultaneous increase in resistance and decrease in force. (d) AFM image of sample after fracture. The arrow indicates the trajectory of the tip between (d) and (e). (e) Image of sample after modification indicated in (d).

As we have shown, breaking the tube, then pushing the ends into each other causes an increase in resistance of only 10%. This makes sense if a large part of the measured resistance comes from the contacts rather than the tube, and thus 13 kΩ may constitute a change on the order of 100% or more in the actual nanotube resistance. We also observe that there was no measurable change in resistance before the tube deformed plastically or fractured, which implies that the change in contact resistance during the modification is negligible, and thus the 15 kΩ change is from the nanotube itself. This is supported by the AFM images of the sample, which shows no change in the contacts, unlike the previously discussed sample. We speculate that this difference may arise from the ratio of the film thickness to the MWNT. If the metal is thick enough, the nanotube is effectively cemented under the lead, and cannot be pulled free.

We have observed the straining and eventual breaking of a nanotube with an AFM probe, while simultaneously measuring the resistance. The strain in the nanotube had no measurable effect on the two-probe resistance, consistent with theoretical work, until the nanotube was strained beyond the elastic limit. The ends of the broken MWNT were pushed into each other and electrical contact between the ends was re-established; though this increased the resistance of the sample by 15 kΩ. We also observed that manipulation of nanotubes has changed the contact resistance by over 100 kΩ. This alerts us that care must be taken in interpreting the contact resistance of nanotube samples as fixed in the presence of external stimuli.

We are grateful to J. Steele, E. Basgall, and B. Davis for help preparing samples. We thank O. Zhou for providing the multi-walled nanotubes and B. Bagnell for SEM characterization of samples. This work was supported by NSF, ONR, NIH/NCRR 02170 and UNC-CH.